\RequirePackage{fix-cm}

\documentclass[smallextended]{svjour3}       
\smartqed  
\usepackage{graphicx}
\newcommand{\Tr}{\mathrm{Tr}}
\newcommand{\tr}{\mathrm{Tr}}

\newcommand{\tI}{\mathrm{I}}
\newcommand{\tII}{\mathrm{II}}

\newcommand{\cC}{\mathcal{C}}
\newcommand{\cD}{\mathcal{D}}

\newcommand{\cI}{\mathcal{I}}
\newcommand{\cJ}{\mathcal{J}}

\newcommand{\be}{\begin{equation}}
\newcommand{\ee}{\end{equation}}
\newcommand{\bea}{\begin{eqnarray}}
\newcommand{\eea}{\end{eqnarray}}
\newcommand{\nn}{\nonumber}
\newcommand{\kt}{\rangle}
\newcommand{\br}{\langle}

\newcommand{\ed}{\end{document}}

\newcommand{\bi}{\begin{itemize}}
\newcommand{\ei}{\end{itemize}}

\newcommand{\bce}{\begin{center}}
\newcommand{\ece}{\end{center}}

\begin{document}

\title{Quantum Teleportation With Nonclassical Correlated States in Noninertial
Frames}

\author{ Hossein Mehri-Dehnavi \and Robabeh Rahimi \and Hosein Mohammadzadeh \and Zahra Ebadi  \and Behrouz Mirza
        }


\institute{H. Mehri-Dehnavi \at
             Department of Physics, Babol University of Technology, Babol, 47148-71167, Iran
           \and
           R. Rahimi \at
             Institute for Quantum Computing, University of Waterloo, Waterloo, ON, N2L 3G1, Canada\\Department of Physics, University of Waterloo, Waterloo, ON, N2L 3G1, Canada\\Department of Chemistry and Molecular Materials Science, Osaka City University, Osaka, 558-8585, Japan
             \email{rrahimid@uwaterloo.ca}
 \and
           H. Mohammadzadeh \at
             Department of Physics, University of Mohaghegh Ardabili,  P.O. Box 179, Ardabil, Iran
 \and
           Z. Ebadi, B. Mirza \at
             Department of Physics, Isfahan University of Technology, Isfahan, 84156-83111, Iran
}

\date{Received: date / Accepted: date}

\maketitle

\begin{abstract}

Quantum teleportation  is studied in noninertial frame, for
fermionic case, when Alice and Bob share a general nonclassical
correlated state. In noninertial frames two fidelities of
teleportation are given. It is found that the average fidelity of
teleportation from a separable and nonclassical correlated state is
increasing with the amount of nonclassical correlation of the state.
However, for any particular nonclassical correlated state, the
fidelity of teleportation decreases by increasing the acceleration.
\end{abstract}

\section{Introduction}
Quantum teleportation, initially proposed by Bennett \textit{et.
al.} \cite{n1}, is one of the important  quantum protocols because
of its several theoretical features and interesting applications.
Quantum teleportation is the reliable transfer of quantum state by
using a shared source of entanglement, in addition to a classical
communication channel.

The original teleportation assumes Alice and Bob are sharing a
perfect bipartite entangled pair of particles. An unknown quantum
state is supposed to be teleported from Alice to Bob. Alice measures
the unknown state and the part of the shared entanglement in her
disposal, in Bell basis, and sends off the outcome of the
measurement, in the form of two bits of classical information, to
Bob. Consequently, Bob, upon receiving the classical information,
applies appropriate unitary operations and transforms the quantum
state to the original one that Alice had.

Since its original proposal, quantum teleportation  has been relying
on the concept of quantum entanglement. Quantum entanglement
initially appeared as a source of paradoxical features of quantum
mechanics \cite{EPR}.  There are several measures for quantum entanglement. Logarithmic
Negativity \cite{negativity} is a measure of entanglement of
bipartite
 states, and is defined as
    \bea
    N(\rho):
    =\log_2\sum_i|\lambda_i(\rho^{\rm pt})|,
    \label{logneg}
    \eea
where, $\lambda_i(\rho^{\rm pt})$ denotes the eigenvalues of the
partial transpose, $\rho^{\rm pt}$, of the density matrix  $\rho$ of
a bipartite quantum system $AB$.

In the lack of a mathematical proof for the distinct and unique role
of entanglement for quantum information processing and quantum
computing, in general, existence of any nonclassical correlation has
been candidated for the expected super-power source for quantum
processors \cite{datta}. Deterministic Quantum Computation with One Quantum Bit (DQC1)
\cite{Knill} that contains very little or no bipartite entanglement,
performs a computation that has no known efficient classical
algorithm.

According to Oppenheim-Horodecki paradigm \cite{ref31}, a
nonclassical correlated state is a state that cannot be represented
in the form of a ``properly classically correlated state'', $
\rho_{\rm pcc}$, with the following definition
    \be
    \rho_{\rm pcc}=\sum_{i=1}^{d^A}\sum_{j=1}^{d^B}e_{ij}
    |v_A^i\kt\br v_A^i|\otimes|v^j_B\kt\br v^j_B|,
    \label{properly-CC}
    \ee
where, $d^A$ and $d^B$ are the dimensions of the Hilbert spaces of
$A$ and $B$, respectively, and $e_{ij}$ is the eigenvalue of
$\rho_{\rm pcc}$ corresponding to an eigenvector
$|v_A^i\kt\otimes|v^j_B\kt$. Quantum Discord, $\cD$, is one of the
most studied measures for nonclassical correlation \cite{pra2008}.
It is defined \cite{ollivier} as the discrepancy between the quantum
mutual information, $\cI$, and the locally accessible mutual
information, $\cC$,
    \bea
    \cD(A:B)=\cI(A:B)-\cC(A:B),
    \label{discord}
    \eea
with the following definitions
    \bea
    \cI(A:B)=S(\rho_{A})+S(\rho_{B})-S(\rho),
    \label{mutual}
    \eea
    and
    \bea
    \cC(A:B)=\max_{\{\Pi_k\}} \left[ \cJ_{\{\Pi_k\}}(A:B)\right],
    \label{CC}
    \eea
where, $\rho_A$ and $\rho_B$ are the reduced density operators of
$A$ and $B$, respectively. $S(\rho)=-\tr(\rho\log_2\rho)$ is the von
Neumann entropy.  $\{\Pi_k\}$'s are von Neumann operators acting on
subsystem $B$ and corresponding to the outcome $k$.
 $\cJ$ is locally accessible mutual information defined as
    \bea
    \cJ_{\{\Pi_k\}}(A:B)=S(\rho_{A})-S_{\{\Pi_k\}}(A|B),
    \label{J}
   \eea
 $S_{\{\Pi_k\}}(A|B)$ is the quantum conditional entropy,
defined as
    \bea
    S_{\{\Pi_k\}}(A|B)=\sum_k p_k S(\rho_{A|k}),
    \label{S(A:B)}
    \eea
where, $\rho_{A|k}=\tr_B(\Pi_k \rho \Pi_k)/p_k$, with $p_k=\tr(\Pi_k
\rho \Pi_k)$.

Quantum teleportation by using not completely entangled state has
been studied \cite{arxiv1}. Also, it has been shown that a separable
state which involves nonclassical correlation can be used for
quantum information transmission \cite{arxiv2}. Extension of quantum
teleportation to noninertial frames has been perviously studied
 in an approach different than in our paper
\cite{Alsing-prl}. In fact, the previously studied process should be
regarded as a ``noninertial frame observation'' of quantum
teleportation   since the involved  entangled state in the
teleportation is not appropriately affected by changing to a
noninertial frame.  In this work, we keep the original teleportation
but apply the appropriate changes on every steps, accordingly.

Suppose that Alice, $A$, is resting and Rob, $R$, is the uniformly
accelerating Bob, with the acceleration $a$. The corresponding
Minkowski spinor basis states
 \cite{Alsing-prl,annals}
 are as follows
    \bea
    |0\kt_M&=&\cos r |0\kt_\tI|0\kt_{\tII}+\sin r |1\kt_\tI|1\kt_{\tII},
    \label{0-m-to-r}\\
    |1\kt_M&=&|1\kt_\tI|0\kt_{\tII},
    \label{1-m-to-r}
    \eea
where, $\cos r=1/\sqrt{1+e^{-2\pi \omega c/a}}$ and
$\omega=\mbox{$\sqrt{|\vec{k}|^2+m^2}$}$ denotes the energy of any
mode with momentum $\vec{k}$ and mass $m$. Here, the subscripts
$\tI$ and $\tII$ represent Rindler regions $\tI$ and $\tII$ Fock
states, respectively.

Quantum teleportation has been demonstrated in different physical
systems,
 including NMR \cite{BB}. In NMR  and other similar
bulk ensemble quantum computation such as electron nuclear double
resonance (ENDOR) \cite{ref23}, practically a pseudo-entanglement of
the following form is generated
    \bea
    \rho_{\rm pe}=\frac{1-p}{4}I+p|\Phi^+\kt\br\Phi^+|,
    \label{pseu-ent-gen}
    \eea
when intentionally the entangled state
$|\Phi^+\kt=\frac{1}{\sqrt{2}}(|00\kt+|11\kt)$ is the desired
 state. Here $\rho_{\rm pe}$ is entangled  if  the purity $p>1/3$.
Generally, experimental conditions bring down the state to a region
where entanglement is absent and the workable state is just a
nonclassical correlated state, Eq. (\ref{pseu-ent-gen}). Detecting
the status of entanglement or nonclassical correlation of the
involved states for these physical systems has been practically of
importance \cite{rs-pra2010}. It should be noted also that the
quantum  teleportation implemented with such physical systems may
not absolutely be relying on a pure entangled state. In this work,
we study quantum teleportation with a general nonclassical
correlated state, Eq. (\ref{pseu-ent-gen}), so the results will be
applicable to the above mentioned physical systems.

This paper is organized as follows. In next section, the conventional
quantum teleportation with maximal entanglement is described when an
observer in noninertial frame is detecting the resultant teleported
state. In  section 3, quantum teleportation is generalized for the
case that a nonclassically correlated state is used, and Rob, the
accelerated Bob,  is in a noninertial frame, so his preshared
nonclassical correlated state is also affected, accordingly.   The
paper will be concluded by bringing discussions in the last section.

\section{A noninertial frame observation of quantum teleportation with maximally entangled state}
Consider an arbitrary one-qubit state
$|\psi\kt=\alpha|0\kt+\beta|1\kt$, which Alice wishes to teleport to
Bob. If Alice and Bob have preliminarily shared the bipartite state
$|\Phi^+\kt$, the total initial state is given by $|\Psi_{\rm
in}\kt= |\psi\kt |\Phi^+\kt$. The  first two qubits are in Alice's
possession and the third one belongs to Bob. Alice  starts quantum
teleportation by applying CNot and Hadamard (H) gates, thus changes
the total state to $|\Psi'\kt=({\rm H}\otimes 1 \otimes 1)({\rm
CNot}\otimes 1) |\Psi_{\rm in}\kt$. Alice measures the state in her
possession, in $Z$-basis, and extracts the state $|ij\kt$,
$i,j=\{0,1\}$. The total state is then $|ij\kt|\phi_{ij}\kt$ with
probability of $p_{ij}=\Tr (\br ij|\Psi'\kt\br \Psi'|ij\kt )$, where
$|\phi_{ij}\kt=X^jZ^i|\psi\kt$. Alice sends the results of the
measurement, $i$ and $j$, to Bob using classical information
channels. Bob retrieves the state, supposed to be teleported,  by
performing the quantum gate $(X^jZ^i)^{-1}=Z^iX^j$  on the qubit in
his possession. The result is given by
$|\tilde{\psi}\kt=Z^iX^j|\phi_{ij}\kt=|\psi\kt$, and the
corresponding teleportation fidelity is  $F=|\br\tilde{ \psi}
|\psi\kt|^2=1$.

In \cite{Alsing-prl}, Alice finds the values $i$, $j$, and sends
them to Bob.  Bob rewrites the state $|\phi_{ij}\kt$ in Rindler
frame by using Eqs. (\ref{0-m-to-r}-\ref{1-m-to-r}) to find the Rob
and anti-Rob states, $|\phi^{ij}_{\tI, \tII}\kt$. By tracing out the
anti-Rob, $\tII$, modes, he  finds the Rob density matrix,
$\rho^{ij}_{\tI}$. Finally, by applying the operator $Z^iX^j$ on the
density matrix, he finds
${\tilde{\rho}}^{ij}_{\tI}=Z^iX^j \rho^{ij}_{\tI} (Z^iX^j)^{-1}$.

It is clear from the above notation that, in general, the state
${\tilde{\rho}}^{ij}_{\tI}$ depends on Alice's measurement results,
$i$ and $j$. This fact comes from the non-symmetric property of
transformations, Eqs. (\ref{0-m-to-r}-\ref{1-m-to-r}), for $|0\kt$
and $|1\kt$. In order to make the result ${\tilde{\rho}}^{ij}_{\tI}$
independent of the values $i$ and $j$, the symmetric dual-rail basis
set is used. The indexes are omitted and the teleported state is
written as ${\tilde{\rho}}_{\tI}={\tilde{\rho}}^{ij}_{\tI}$. The
fidelity of teleportation  is given by $F=\br \psi
|{\tilde{\rho}}_{\tI}|\psi\kt=\cos ^2 r$, for fermionic case
\cite{Alsing-prl}.

There are  objections to this study. Quantum teleportation is
studied in noninertial frame with a cost that the original
teleportation protocol is modified. Recalling the original
teleportation, Bob is not only an observer but he is the party who
receives the classical information and applies accordingly changes
to the entangled part in his disposal to extract the desired
teleportation state. This implies that if Bob is assumed in the
accelerated frame, so is called Rob, then his belonging pre-shared
entanglement also should be modified according to the acceleration.
Also, the original teleportation protocol is not restricted to any basis set,
and is universal. Therefore, even extending quantum teleportation to
noninertial frame should be practically possible without any
restriction such as dual-rail basis set.

 In following, we extend the original quantum
teleportation to noninertial frame, in addition, we generalize our
study by using a nonclassical channel of the form Eq.
(\ref{pseu-ent-gen}) instead of a pure entangled state.

\section{Teleportation with nonclassical correlated state in noninertial frames}

 Alice wants to teleport the state
$|\psi\kt$ to Bob, by using the shared state $\rho_{\rm pe}$. The
initial state is given by ${\varrho}_{\rm in}=
|\psi\kt\br\psi|\otimes \rho_{\rm pe}$, where the first two qubits
of ${\varrho}_{\rm in}$ are the Alice's ones and the third one is
the Bob's qubit. Substituting $p=1$ gives the special case of
teleportation with maximally entangled state $|\Phi^+\kt$. Here, we
shall recall Rob, the uniformly accelerated Bob, following the
general convention.

If Rob's state starts to degrade with constant acceleration $a$ then
the bipartite  state, $\rho_{\rm pe}$, transforms to a tripartite
state $\rho_{A,\tI,\tII}$, using Eqs.
(\ref{0-m-to-r}-\ref{1-m-to-r}). $\rho_{A,\tI,\tII}$ is the quantum
state of Alice (A), Rob (I), and anti-Rob (II). Tracing out the
anti-Rob modes results the Alice-Rob density matrix,
$\rho_{A,\tI}=\tr_{\tII} (\rho_{A,\tI,\tII})$ as follows
\cite{annals}
     \bea
      \rho_{A,\tI}=\frac{1}{4}\left(
                          \begin{array}{cccc}
                            (1+p)\cos^2 r & 0 &  0& 2 p\cos r  \\
                            0 & 1+\sin^2 r-p\cos^2 r & 0 & 0 \\
                            0 & 0 & (1-p)\cos^2r & 0 \\
                            2 p\cos r& 0 & 0 & 1+\sin^2 r+p\cos^2 r \\
                          \end{array}
                        \right).\nn
    \eea
 We have  used the basis $|0\kt_A|0\kt_{\tI}$,
$|0\kt_A|1\kt_{\tI}$, $|1\kt_A|0\kt_{\tI}$, and $|1\kt_A|1\kt_{\tI}$
to write the above matrix. Therefore, we should use
${\varrho}_{A,\tI}= |\psi\kt\br\psi|\otimes \rho_{A,\tI}$ instead of
the initial state ${\varrho}_{\rm in}$, for teleportation with
uniformly accelerated partner, Rob.

Alice starts the  teleportation procedure by performing CNot and
Hadamard gates on the particles in her possession. Then, she
measures  the state  in $Z$-basis. The total state collapses to
$|ij\kt\br ij|\otimes {\rho}^{ij}_{\tI}$, with probability
$p_{ij}=\Tr (\br ij|{\varrho'}_{A,\tI}|ij\kt )=1/4$, where
${\rho}^{ij}_{\tI}$ are given by
\bea
      {{\rho}}^{i0}_{\tI}=\frac{1}{2}\left(
                          \begin{array}{cc}
                            [1+p(|\alpha|^2-|\beta|^2)]{\cos^2 r} &2(-1)^{i} p\alpha \beta^* \cos r \\
                            2(-1)^{i}p\alpha^* \beta \cos r& 1+{\sin^2 r}-p(|\alpha|^2-|\beta|^2)\cos^2r \\
                          \end{array}
                        \right), \label{rho-I-i0}
  \\
      {{\rho}}^{i1}_{\tI}=\frac{1}{2}\left(
                          \begin{array}{cc}
                            [1-p(|\alpha|^2-|\beta|^2)]{\cos^2 r} & 2(-1)^{i}p\alpha^* \beta \cos r\\
                            2(-1)^{i} p\alpha \beta^* \cos r& 1+{\sin^2 r}+p(|\alpha|^2-|\beta|^2)\cos^2r \\
                          \end{array}
                        \right).
    \label{rho-I-i1}
    \eea
Alice sends the results of the measurement, $i$ and $j$,  to Rob by
using classical information channels. Consequently, Rob extracts
information by performing  the quantum gate $(X^jZ^i)^{-1}=Z^iX^j$
on the state in his hand, ${\tilde{\rho}}^{ij}_{\tI}=Z^iX^j
\rho^{ij}_{\tI} (Z^iX^j)^{-1}$, as follows
  \bea
     && {\tilde{\rho}}^{i0}_{\tI}= {\rho}^{00}_{\tI},
    \\ && {\tilde{\rho}}^{i1}_{\tI}=\frac{1}{2}\left(
                          \begin{array}{cc}
                             1+{\sin^2 r}+p(|\alpha|^2-|\beta|^2)\cos^2r& 2p\alpha \beta^* \cos r \\
                            2p\alpha^* \beta \cos r& [1-p(|\alpha|^2-|\beta|^2)]\cos^2 r \\
                          \end{array}
                        \right),\nn\\
    \label{rho-I-till-ij}
    \eea

where ${\rho}^{00}_{\tI}$, is given by Eq. (\ref{rho-I-i0}).

 For  $p=1$, $r\neq0$,  we have
quantum teleportation with maximally entangled state in noninertial
frame. The result is different than \cite{Alsing-prl} because we use
a general non-symmetric basis set instead of the symmetric dual-rail
basis. Also, if $r=0$, for a general $p$, the quantum teleportation
is same as to the one in inertial frame using a general nonclassical
correlated state. The resultant state extracted  by Rob is
independent of the results of the Alice's measurement, $i$ and $j$,
as it is expected.

For a general case  the fidelities, $F_{ij}=\br \psi
|{\tilde{\rho}}^{ij}_{\tI}|\psi\kt$, are  given as follows
    \bea
     F_{i0}&=&\frac{1}{2}\left\{ |\beta|^4[2-(1-p)\cos^2r]+|\alpha|^4(1+p)\cos^2 r \right.\nn\\
     &&\left.+2|\alpha|^2|\beta|^2[p(2-\cos
     r)\cos r+1]\right\},\label{Fij0}\\
     F_{i1}&=& F_{i0}+(|\alpha|^4-|\beta|^4)\sin^2 r.
     \label{Fij}
     \eea
{We evaluate the average fidelity by using the Bloch sphere
representation of the initial state, namely
$|\psi\kt=\alpha|0\kt+\beta|1\kt:=\cos
\frac{\theta}{2}|0\kt+e^{i\phi}\sin \frac{\theta}{2}|1\kt$. Thus,
the average fidelity is given by,
 \be
    <F_{ij}>=\frac{1}{4\pi}\int_0^{2\pi}\int_0^{\pi}F_{ij}(\theta, \phi)
    \sin \theta d\theta d \phi.
  \label{Fij-av}
 \ee
Using equations (\ref{Fij0}-\ref{Fij})  the average fidelity is
obtained as follows,
    \bea
    <F>=<F_{ij}>=\frac{1}{6}\left( 3+p\cos^2r+2 p\cos
    r \right).
  \label{F-av}
 \eea

Logarithmic negativity and discord for $\rho_{A,\tI}$ should be
calculated in order to evaluate the contributions from entanglement
and nonclassical correlation to the above fidelity of teleportation.
We employ the corresponding calculation results from our previous
work \cite{annals}, where  we have studied the logarithmic
negativity and discord for $\rho_{\rm pe}$ in noninertial frames. In
addition, here, the eigenvalues of the partial transpose of
$\rho_{A,\tI}$ are given in order to study the contribution into the
fidelity of teleportation from a state with a purity of threshold,
$p_{\rm th}$. The quantum state $\rho_{A,\tI}$ with $p_{\rm th}$
involves maximum nonclassical correlation for a separable state.
Hence, the corresponding fidelity of teleportation can be regarded
as the maximum attainable fidelity of teleportation if the
pre-shared state in teleportation is not entangled but involves
nonclassical correlation.
     \bea
    &&\lambda_{1,2}(\rho_{A,\tI}^{\rm pt})=\frac{1}{4}
    \left\{1-p\cos^2r\pm\sqrt{\sin^4r+4p^2\cos^2r}\right\},\nn\\
    &&\lambda_{3,4}(\rho_{A,\tI}^{\rm
    pt})=\frac{1}{4}\left\{1+p\cos^2r\pm\sin^2r\right\}.
    \label{lambda-t-AI}
    \eea
The entanglement threshold for $\rho_{A,\tI}$ is corresponding to
the purity threshold given as
    \bea
         p_{\rm  th}=\frac{3-\cos 2r}{7-\cos 2r}.
         \label{ther-purity-X}
    \eea
As an example $p_{\rm th}=1/3$ for $r=0$. Then by substituting
$p_{\rm th}$ as function of $r$ in Eq.  (\ref{F-av}), we find the
attainable fidelity of teleportation for a state with $p_{\rm th}$,
as plotted in Fig. \ref{fig0}. In this figure, the given fidelity
for any $r$ should be also regarded as the maximum attainable
fidelity of teleportation from a sperable but nonclassically
correlated state. It is clear that the achievable fidelity of
teleportation from a separable but nonclassically correlated state
is a decreasing function of $r$. It has the maximum value $2/3$ for
$r=0$, in accordance with Ref. \cite{Horodecki99}. Also,
teleportation with a separable and classical correlated state gives
fidelity of $1/2$ that is same as the success
 probability from a random guess. Then we
conclude that fidelity of teleportation for a separable and
nonclassical correlated state $<F_{\rm ncc}>$ satisfies $\frac{1}{2}
\leq <F_{\rm ncc}> \leq \frac{2}{3}$, in any noninertial frame, and
the maximum value is achieved for $r=0$, an inertial frame.
\begin{figure}[t]
    \begin{center}
   \includegraphics[scale=.81,clip]{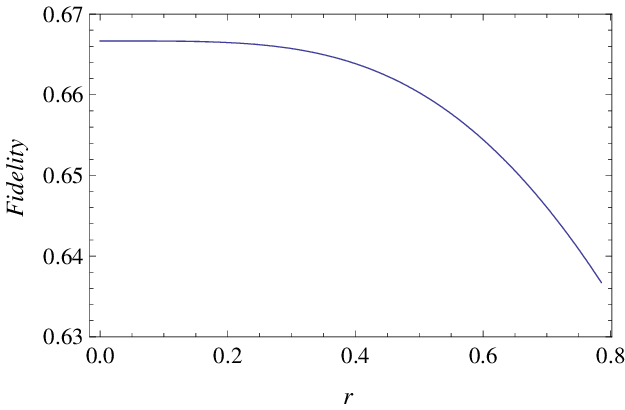}
    {\caption{ The fidelity of teleportation
for a nonclassical correlated state, with the threshold purity, Eq.
(\ref{ther-purity-X}), in noninertial frame with acceleration
corresponding to $r$.
    \label{fig0}}}\end{center}
    \end{figure}

\begin{figure}[t]
    \begin{center}
   \includegraphics[scale=.81,clip]{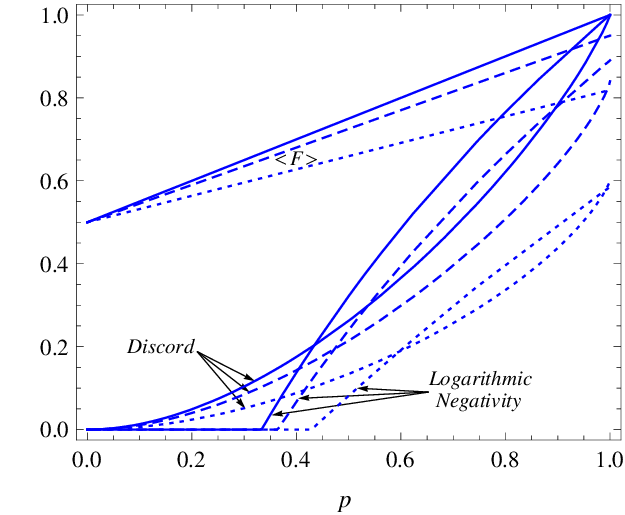}
    {\caption{ For the extreme cases $r=0$
(the solid lines), and $r=\pi/4$ (the dotted lines),
    in addition to an intermediate case $r=\pi/8$ (the dashed lines), the average fidelity of
     teleportation, $<F>$, discord, and logarithmic negativity are plotted as functions of the purity, $p$.
    \label{fig2}}}\end{center}
    \end{figure}

\section{Discussions and conclusion}
In this work, we studied quantum teleportation with nonclassical
correlated state in noninertial frame. Fidelity of teleportation,
discord, and logarithmic negativity  are evaluated as functions of
$r$, corresponding to the acceleration $a$, and the purity, $p$, of
the state.

 Fig. \ref{fig2} shows the fidelity of teleportation, discord, and
the logarithmic negativity for two extreme accelerations,
corresponding to $r=0$ and $r=\pi/4$, in addition to an intermediate
case, $r=\pi/8$. The logarithmic negativity for $r=0$ is nonzero for
$p>p_{\rm th}=1/3$ and increases to reach the maximum $1$, that is
when the state, Eq. (\ref{pseu-ent-gen}), is a Bell state. For
$r=\pi/8$, $p_{\rm th}$ is $0.364$,  and for $r=\pi/4$, $p_{\rm th}$
is $3/7$, and the logarithmic negativity is an increasing function
with the maximum values $N(p=1,r=\pi/8)=0.890$,
$N(p=1,r=\pi/4)=0.585$. In Fig. \ref{fig2}, discord is nonzero for
any nonzero purity, $p$, and it has maximum value same as the values
for the corresponding logarithmic negativity. In this figure, the
fidelity of teleportation is an increasing function with the purity.
It is $1/2$ only for a separable and a classical correlated state.
Any nonclassical correlation is sufficient for extracting fidelity
of teleportation greater than $1/2$. Specifically, for $r=0$, the
fidelity of teleportation from a separable and nonclassical
correlated state calculated from this work is in a good agreement
with the original work by Horecki et al. \cite{Horodecki99}, in
which the optimal fidelity of teleportation in an inertial frame,
$r=0$, is given as a function of the maximally attainable singlet
function. To be more precise in this circumstance, the optimal
fidelity is calculated to be $2/3$ for a noisy singlet Eq. (13) of
\cite{Horodecki99}, as a generalization of the $2 \times 2$ Werner
state. This fact is generally studied for any $r$ and the results
are illustrated in Fig. \ref{fig3}.

\begin{figure}[t]
    \begin{center}
   \includegraphics[scale=.81,clip]{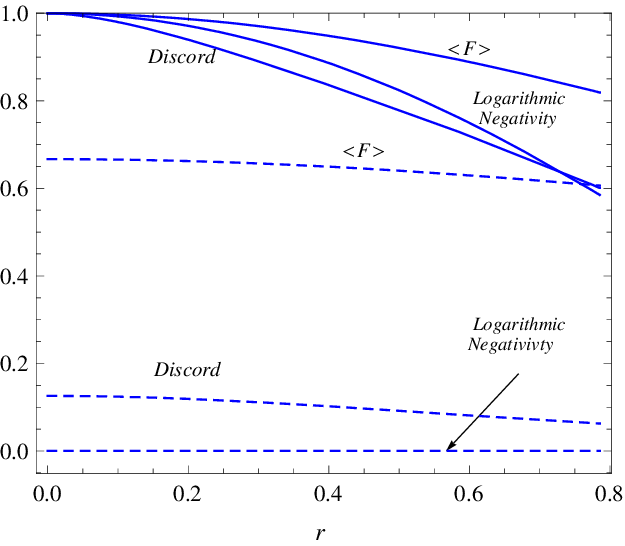}
    {\caption{The average fidelity, $<F>$,
discord and logarithmic
    negativity, as functions of $r$.
     The upper-full curves are plotted for the maximally entangled states, $p=1$, and lower-dashed curves are plotted for   $p=1/3$.
    \label{fig3}}}\end{center}
    \end{figure}
\begin{figure}
 \end{figure}

The special cases, $p=1/3$ and $p=1$ are shown in Fig. \ref{fig3}.
Logarithmic negativity is zero for $p=1/3$, regardless of the
acceleration. However, $\cD (p=1/3, r=0)=0.126$, $<F(p=1/3,
r=0)>=2/3$, and these functions decrease by increasing $r$, and
reach the minimum values $\cD(p=1/3, r=\pi/4)=0.063$,
$<F(p=1/3,r=\pi/4)>=0.606$. For $p=1$, all the three functions start
from the maximum value one, and decrease to different minimum
values, $N(p=1, r=\pi/4)=0.585$, $\cD (p=1, r=\pi/4)=0.601$, and
$<F(p=1, r=\pi/4)>=0.819$.

Any nonzero nonclassical correlation gives fidelity of teleportation
larger than the achievable fidelity from a purely classical state,
and the fidelity of teleportation is generally decreasing with
increasing acceleration in noninertial frame.

\begin{acknowledgements}
We would like to acknowledge Animesh Datta, Gerardo Adesso, and Laleh Memarzadeh for useful
comments and discussions. RR would like to acknowledge supports from
Industry of Canada, CIFAR and NSERC.
\end{acknowledgements}


\begin{thebibliography}{}

\bibitem{n1}
Bennett, C. H., Brassard, G., Cr\'{e}peau, C., Jozsa, R., Peres, A.,
Wootters, W. K.: Teleporting an unknown quantum state via dual
classical and Einstein-Podolsky-Rosen channels, Phys. Rev. Lett.
{\bf 70}, 1895 (1993).

\bibitem{EPR}
Einstein, A., Podolsky, B., Rosen, N.: Can quantum-mechanical
description of physical reality be considered complete?, Phys. Rev.
{\bf 47}, 777 (1935).

\bibitem{negativity}
 Peres, A.: Separability criterion for density matrices,  Phys. Rev. Lett. {\bf 77}, 1413 (1996);\\ \.{Z}yczkowski, K., Horodecki, P.,
Horodecki, A., Horodecki, M.: Volume of the set of separable states, Phys. Rev. A {\bf 58}, 883 (1998).

\bibitem{datta}
 Datta, A.,  Shaji, A., Caves, C. M.: Quantum discord and the power of one qubit, Phys. Rev. Lett. {\bf 100}, 050502 (2008).

\bibitem{Knill}
 Caves, E., Laflamme, R.: Power of one bit of quantum information, Phys. Rev. Lett. {\bf 81}, 5672 (1998).

\bibitem{ref31}
Horodecki, M., Horodecki, P., Horodecki, R.,  Oppenheim, J., Sen(De), A., Sen, U., Synak-Radtke, B.: Local versus nonlocal information in quantum-information theory: Formalism and phenomena, Phys. Rev. A {\bf 71}, 062307 (2005).

\bibitem{pra2008}
SaiToh, A.,  Rahimi, R.,   Nakahara, M.: Nonclassical correlation in a multipartite quantum system: Two measures and evaluation, Phys. Rev. A {\bf 77}, 052101 (2008); SaiToh, A.,  Rahimi, R., Nakahara, M.: Evaluating measures of nonclassical correlation in a multipartite quantum system,  Int. J. Quantum Inform. {\bf 06}, 787 (2008).

\bibitem{ollivier}
 Ollivier, H., Zurek, W. H.: Quantum Discord: A measure of the quantumness of correlations, Phys. Rev. Lett. {\bf 88}, 017901 (2001).

\bibitem{arxiv1}
 Zhao, M. J.,  Li, Z. G.,  Fei, Sh.,  Wang, Z. X., Li-Jost, X.: Faithful teleportation with arbitrary pure or mixed resource states,  J.
Phys. A: Math. Theor. {\bf 44}, 215302 (2011).

\bibitem{arxiv2}
 Wang, L., Huang, J.,  Dowling, J. P.,  Zhu, Sh.: Quantum information transmission, Quantum Inf. Process. {\bf 12}, 899
 (2013).

\bibitem{Alsing-prl}
 Alsing, P. M.,  Milburn, G. J.: Teleportation with a uniformly accelerated partner, Phys. Rev. Lett. {\bf 91}, 180404 (2003);\\ P.
 Alsing, P. M., McMahon,  D.,  Milburn, G. J.: Teleportation in a non-inertial frame, J. Opt. B {\bf 6}, S834 (2004).

\bibitem{annals}
 Mehri-Dehnavi, H., Mirza, B., Mohammadzadeh, H.,   Rahimi, R.: Pseudo-entanglement evaluated in noninertial  frames,
Ann. Phys. {\bf 326},  1320 (2011).

\bibitem{BB}
Nielsen, M. A.,  Knill,  E.,  Laflamme, R.: Complete quantum teleportation using nuclear magnetic resonance, Nature {\bf 396}, 52 (1998).

\bibitem{ref23}
Rahimi, R.,  Sato, K., Furukawa, K.,  Toyota,  K., Shiomi, D., Nakamura, T.,  Kitagawa, M.,  Takui,  T.: Pulsed ENDOR-based quantum information processing, Int. J. Quantum Inf. {\bf 3}, 197 (2005).


\bibitem{rs-pra2010}
 Rahimi, R., Takeda, K.,   Ozawa, M.,  Kitagawa,  M.: Entanglement witness derived from NMR superdense coding, J. Phys. A: Math. Gen. {\bf 39}, 2151 (2006); Rahimi, R., SaiToh, A., Nakahara, M.,  Kitagawa,  M.: Single-experiment-detectable multipartite entanglement witness for ensemble quantum computing, Phys. Rev. A {\bf 75}, 032317 (2007); Rahimi, R., SaiToh, A.: Single-experiment-detectable nonclassical correlation witness, Phys. Rev. A {\bf 82}, 022314 (2010).

\bibitem{Horodecki99}
Horodecki, M.,  Horodecki, P.,  Horodecki,  R.: General teleportation channel, singlet fraction, and quasidistillation,
Phys. Rev. A  {\bf 60}, 1888 (1999).

\end{thebibliography}
\end{document}